\documentclass[11pt,twoside]{article}  
\usepackage{asp2006}
\usepackage{adassconf}

\begin{document}   

\paperID{A.16}

\title{Non-linear Least Squares Fitting in IDL with MPFIT}

\markboth{Markwardt}{Non-Linear Fitting with MPFIT}

\author{Craig B. Markwardt\altaffilmark{1,2}}
\altaffiltext{1}{Department of Astronomy, 
                 University of Maryland, College Park, MD, USA}
\altaffiltext{2}{CRESST and Astroparticle Physics Laboratory, 
                 NASA/GSFC, Greenbelt, MD, USA}

\contact{Craig Markwardt}
\email{Craig.Markwardt@nasa.gov}

\paindex{Markwardt, C.~B.}



\begin{abstract}          
MPFIT is a port to IDL of the non-linear least squares fitting
program MINPACK-1.  MPFIT inherits the robustness of the
original FORTRAN version of MINPACK-1, but is optimized for
performance and convenience in IDL.  In addition to the main fitting
engine, MPFIT, several specialized functions are provided to fit 1-D
curves and 2-D images; 1-D and 2-D peaks; and interactive fitting from
the IDL command line.  Several constraints can be applied to model
parameters, including fixed constraints, simple bounding constraints, and
``tying'' the value to another parameter.  Several data weighting
methods are allowed, and the parameter covariance matrix is computed.
Extensive diagnostic capabilities are available during the fit, via a
call-back subroutine, and after the fit is complete.  Several
different forms of documentation are provided, including a tutorial,
reference pages, and frequently asked questions.  The package has been
translated to C and Python as well.  The full IDL and C packages can 
be found at \htmladdURL{http://purl.com/net/mpfit}.
\end{abstract}

\section{Introduction\label{sec:intro}}
Non-linear least squares fitting is an integral part of most
astronomical analysis.  The process embodies the fundamental process
of hypothesis testing for a candidate model which may explain the
data.  There are several built-in fitting procedures packaged within
the Interactive Data Language (IDL) product\footnote{IDL is a product
of ITT Visual Information Solutions, \htmladdURL{http://ittvis.com/}}.
Unfortunately, the existing IDL procedures are not very desirable
from the perspective of astronomical data analysis.  The built-in
procedures CURVEFIT and LMFIT are somewhat unreliable, and do not
always take advantage of IDL's vectorization capability.  Because of
these limitations, the author undertook to write a robust and
functional least squares fitting code for IDL.  The work was based on
translating the highly successful MINPACK-1 package written in FORTRAN
into IDL, and building new functionality upon that framework.

\section{The Heritage of MINPACK\label{sec:minpack}}
MPFIT is basically a translation and enhancement of the MINPACK-1
software, originally developed by Jose Mor\'{e} collaborators at
Argonne National Laboratories.  The code was written in FORTRAN, and
is available now from the NETLIB software repository.  MINPACK-1 has
the advantages that it is:
\begin{itemize}
\item robust --- designed by numerical analysts with real data in mind
\item self-contained --- not dependent on a large external library
\item general --- capable of solving most non-linear equations
\item well-known --- one of the most-used libraries in optimization problems
\end{itemize}
The original MINPACK-1 library contains two different versions, {\tt
lmder} and {\tt lmdif}.  Both require the user function to compute the
residual vector, $r$, but {\tt lmder} also requires the user to
compute the Jacobian matrix, $J$, of the residuals as well; {\tt
lmdif} estimates the Jacobian via finite differences.  The MINPACK
algorithm solves the problem by linearizing it around the trial
parameter set, $p_o$, and solving for an improved parameter set, $p =
p_o + \delta p$, via the least squares equation, $(J^T J) \delta p = -
J^T r.$ The solution is obtained by $QR$ factorization of $J$, leading
to improved numerical accuracy over the normal equations form.  The
standard Levenberg-Marquardt technique of replacing the first
parenthesized term with $(J^T J + \lambda D^T D)$, where $\lambda$ is
the Levenberg Marquardt parameter and $D$ is a diagonal scaling
matrix, produces faster convergence.  The solution is iterated until
user-selected convergence criteria are achieved, based on the sum of
squares and residual values.

\section{Translation to IDL\label{sec:implementation}}

The translation to IDL focused on preserving the quality of the
original code, optimizing it for speed within IDL, and adding
functionality within the semantics of IDL.  The result of the
translation is a single fitting engine, {\tt MPFIT}, which provides
all of the original MINPACK-1 capability.  This function is not
specific to a particular problem, i.e. it can be used on data of
arbitrary dimension or weighting.

In addition to the generic fitting routines, several convenience
routines have been developed that make {\tt MPFIT} useful in several
specific problem domains:
\begin{itemize}
\item {\tt MPFITFUN}, {\tt MPFIT2DFUN} --- optimized for 1-D \& 2-D functions;
\item {\tt MPFITEXPR} --- for dynamically-created formulae, e.g. on the command line;
\item {\tt MPCURVEFIT} --- a drop-in replacement for the standard {\tt CURVEFIT} IDL library routine, for users who need compatibility;
\item {\tt MPFITPEAK}, {\tt MPFIT2DPEAK} --- specialized for 1-D \& 2-D peak fitting;
\item {\tt MPFITELLIPSE} --- for fitting elliptical curves to X/Y scatter points.
\end{itemize}

\noindent
The IDL version can be found on the author's website (see Resources, sec.~\ref{sec:resources}).

\section{Innovations of MPFIT\label{sec:innovations}}
Beyond the original MINPACK-1 code, MPFIT contains several innovations
which enhances its usefulness and convenience to the user, and also
take advantage of the capabilities of IDL.

{\bf Private Data.} The user can pass any private data safely to the
user function as keyword variables via the {\tt FUNCTARGS} parameter.
This helps to avoid the use of common block variables.


{\bf Parameter Constraints.}  The notion of simple parameter boundary
constraints is supported via the {\tt PARINFO} parameter.
Individually settable upper and lower limits are supported via {\tt
LIMITS}.  Also, as a convenience, parameters can be held {\tt FIXED},
or {\tt TIED} to another parameter value.  

The total number of degrees of freedom is tracked, as well as the
number of parameters pegged at their limits (via the {\tt DOF} and
{\tt NPEGGED} keywords).

{\bf Jacobian Calculations.}  The user is free to supply explicit
derivatives in their user function, or have MPFIT calculate them
numerically, depending on the {\tt AUTODERIVATIVE} and {\tt
PARINFO.MPSIDE} settings.  The method for calculation of
derivatives (step size and direction) are settable on a per-parameter
basis via the {\tt PARINFO.STEP} and {\tt .RELSTEP} settings.  For
user-calculated derivatives, the user can enable a debugging mode by
setting {\tt PARINFO.MPDERIV\_DEBUG}.

{\bf Covariance matrix.}  The capability to calculate the covariance
matrix of the fit parameters is an improvement over the original
published MINPACK-1 version.

{\bf Hard-to-Compute Functions.}  For functions that are difficult to
compute within a single function call, MPFIT can be requested to allow
`external' evaluation.  MPFIT then returns control temporarily to the
caller so that it can compute the function using external information and 
by whatever means,
and then the caller re-calls {\tt MPFIT} to resume fitting.

{\bf Iteration Function.}  After each iteration, a user procedure
designated by {\tt ITERPROC} may be called.  The default procedure
simply prints the parameter values, but a more advanced version may be
used, for example for GUI feedback.

{\bf Error handling.}  Two error status parameters are provided. Upon
return, {\tt STATUS} is set to a numerical status code suitable for
automated response.  {\tt ERRMSG} is set to a descriptive error string
to inform the human user of the problem.  MPFIT also traps common
problems, like user-function errors and numerical over/under-flows.

\section{Documentation\label{sec:docs}}
MPFIT is provided with extensive documentation.  The MPFIT source
code has reference-style documentation attached to the header of the
source module itself.

A basic tutorial is provided on the author's web page (see
sec.~\ref{sec:resources}), which introduces the user to least squares
fitting of a 1-D data set, and graduates to applying parameter
constraints.  Also, a `FAQ' style web page gives users quick answers
to common questions, such as which module to use, how to calculate
important quantities, and troubleshooting techniques.

\section{Usage\label{sec:usage}}
Examples of usage can be found on the author's website, and as a part
of the code documentation itself.  As an example, consider a user that
has a data set with independent variable {\tt X} and dependent
variable {\tt Y} (with Gaussian errors {\tt EY}), and wants to fit as
a function of {\tt F(X,P)} where {\tt P} is an array of parameters.

In this case, {\tt MPFITFUN} should be used to solve for the best fit
parameters {\tt PBEST} with the following invocation,
\begin{verbatim}

PBEST = MPFITFUN('F', X, Y, EY, PSTART, STATUS=ST, ERRMSG=ERR, $
            BESTNORM=CHI2, DOF=DOF, ERROR=PERROR, COVAR=COVAR)
\end{verbatim}
where {\tt PSTART} is an initial guess of the parameter values.  Upon
return, the best fit $\chi^2$ value and degrees of freedom are
returned via the {\tt BESTNORM} and {\tt DOF} parameters.  Parameter
errors and covariance matrix are returned in the {\tt ERROR} and {\tt
COVAR} parameters.  Error conditions are returned in {\tt STATUS} and
{\tt ERRMSG}.

\section{Results\label{sec:results}}
MPFIT has been available for ten years from the author's web site, and
as been downloaded several thousand times.  During that time, the
package has been continuously improved, both in terms of
functionality, and in terms of fixing ``bugs.''  By its nature, IDL
code is ``open source,'' and at least ten users have contributed
changes which have been incorporated into the main code base.  MPFIT
is distributed with very liberal licensing constraints.

The package has been acknowledged as helpful in a number of published
works, including at least 29 refereed publications since 2001
(including Astrophysical Journal, Monthly Notices and PASJ), and in
102 preprints on the Arxiv preprint server.  Interestingly, MPFIT has
also been translated into the Python language, and is available in the
SciPy scientific package (the interesting aspect is that the
translation was based on the IDL version and not the original
FORTRAN).  The author has also create a C translation of MPFIT, which
has the benefit of speed and portability, along with many of the
IDL-based improvements.

In addition to being used in scientific analysis, MPFIT has also been
incorporated into numerous standalone packages, for example PAN
(``PEAK Analysis'') for neutron scattering spectroscopy, and PintOfAle
for X-ray spectroscopy.

\section{Resources\label{sec:resources}}
\begin{itemize}
\item MPFIT IDL \& C code: 
    \htmladdURL{http://purl.com/net/mpfit}
\item MPFIT Python version:\footnote{Translator: Mark Rivers, U. Chicago} \\
    \htmladdURL{http://cars9.uchicago.edu/software/python/mpfit.html}
\item MINPACK-1 FORTRAN web page:
  \htmladdURL{http://netlib.org/minpack}
\item MINPACK-1 pure C translation: 
  \htmladdURL{http://moshier.net/}
\end{itemize}

\end{document}